\pdfoutput=1

\documentclass[twocolumn,english,superscriptaddress]{revtex4}
\usepackage{times,amssymb,amsmath,graphicx}
\usepackage[bookmarks=false,pdffitwindow=false,pdfstartview={FitH}]{hyperref}
\usepackage[T1]{fontenc}
\usepackage{babel}
\usepackage{color}

\begin{document}


\title{
  Strain-induced Aharonov-Bohm effect at nanoscale and ground state
  of a~carbon nanotube with zigzag edges
}

\author{Adam Rycerz\footnote{Corresponding author; e-mail:
  \href{mailto:rycerz@th.if.uj.edu.pl}{rycerz@th.if.uj.edu.pl}.}}
\affiliation{Institute for Theoretical Physics,
  Jagiellonian University, \L{}ojasiewicza 11, PL--30348 Krak\'{o}w, Poland}

\author{Maciej Fidrysiak}
\affiliation{Institute for Theoretical Physics,
  Jagiellonian University, \L{}ojasiewicza 11, PL--30348 Krak\'{o}w, Poland}

\author{Danuta Goc-Jag\l{}o}
\affiliation{Institute for Theoretical Physics,
  Jagiellonian University, \L{}ojasiewicza 11, PL--30348 Krak\'{o}w, Poland}

\date{September 14, 2023}

\begin{abstract}
Magnetic flux piercing a~carbon nanotube induce periodic gap oscillations
which represent the Aharonov-Bohm effect at nanoscale.
Here we point out, by analyzing numerically the anisotropic Hubbard
model on a~honeycomb lattice, that similar oscillations should be
observable when uniaxial strain is applied to a~nanotube. 
In both cases, a~vector potential (magnetic- or strain-induced) may affect
the measurable quantities at zero field.
The analysis, carried out within the Gutzwiller Approximation, shows
that for small semiconducting nanotube with zigzag edges and realistic value
of the Hubbard repulsion ($U/t_0=1.6$, with $t_0\approx{}2.5\,$eV being
the equilibrium hopping integral) energy gap can be reduced by a~factor
of more than $100$ due to the strain. 
\end{abstract}

\maketitle


\section{Introduction}
Aharonov-Bohm (AB) effect \cite{Sha81,Gef84,Web85} represents one of
the most spectacular features of quantum mechanics, as it demonstrate
the physical meaningfulness of magnetic vector potential when the current
flows entirely through zero magnetic field regions \cite{Naz09}.
In the familiar {\em two-slit-like} setup, conducting region
is pierced by magnetic flux and the quantum interference for an electron
passing simultaneously the two distinct paths, encircling the flux,
is observed.
Since the discovery of graphene \cite{Nov05,Zha05}, several experimental
and theoretical works have addressed specific features of AB effect in
this two-dimensional form of carbon
\cite{Rus08,Sta09,Rec07,Ryc09,Kat10,Kol12,Sch12,Ryc20,Jua11,Pra19}.
In particular, strain-induced pseudomagnetic potentials  allow one to
observe zero magnetic flux analogs of AB effects \cite{Jua11,Pra19}. 

In its nanoscale version realized in carbon nanotubes
\cite{Aji94,Cos04,San11,Shy21},
AB effect no longer requires a~two-slit-like setup because
the flux strongly affects electronic structure near the Fermi energy
\cite{Aji94} (in principle, a~semiconducting nanotube can be turned into
the metallic one and {\em vice versa} \cite{metalfoo}).
This can be traced via measurable quantities for {\em closed}
system rather than by detecting the quantum interference in 
an {\em open} system \cite{Shy21}. 
However, in typical measurements multiwalled nanotubes are used 
\cite{Cos04,San11} and the mutual influence of dynamical and magnetic
phase shifts \cite{Naz09} results in rather complex physics, showing some
common features with quantum interference observed in metallic cylinders 
in Ref.\ \cite{Sha81}.

\begin{figure}[!hb]
  \includegraphics[width=\linewidth]{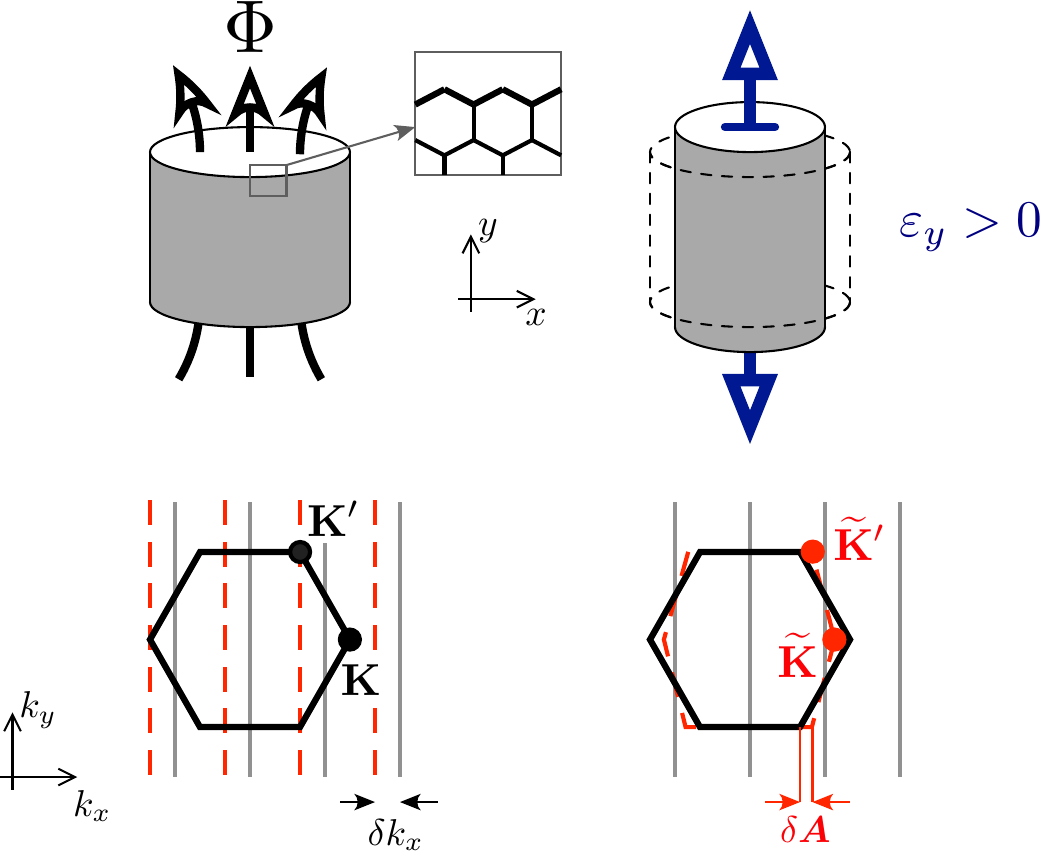}
  \caption{ \label{setup-ab:fig}
    Left: Magnetic flux $\Phi$ parallel to the main axis ($y$) of
    zigzag $(N_x,0)$ nanotube (top) shifts the values of quantized transverse
    momenta, $k_x=2\pi{}n_x/N_x$ with $n_x=0,1,\dots,N_x-1$, 
    by $\delta{}k_x=2\pi(e/h)\,\Phi/N_x$ (bottom).
    Right: In analogy, uniaxial strain $\varepsilon_y>0$ affects the low-energy
    spectrum by shifting the Dirac points, {\boldmath$K$} and {\boldmath$K'$},
    by $\delta{}\mbox{\boldmath$A$}=(\mp{}\delta{}A_x,0)$, with
    $\delta{}A_x\propto{}\varepsilon_y$ [see the main text for details].
    Both mechanisms may lead to approximately periodic bandgap oscillations,
    with the minimum gap corresponding to a~discrete value of $k_x$
    matching the Dirac point. 
  }
\end{figure}

Here, we discuss a~nanoscale AB effect in a~strained nanotube, following
the analogy visualized in Fig.\ \ref{setup-ab:fig}.
The attention is focussed on zigzag nanotube, as the strain along armchair
direction does not open an {\em intrinsic} (i.e., size-independent) gap
in the single-particle spectrum \cite{Per09} allowing also to complement
the discussion on the role of electron correlations in graphene-related
systems \cite{Sor12,Sch13,Zha21,Rut23}.
Our results show that strains earlier demonstrated for planar graphene
samples \cite{McR19,Zhe23}, if accessed in a~single-walled nanotube device,
may result in significant, approximately periodic evolutions of the
multiparticle ground state. For the Hubbard repulsion $U/t_0=1.6$
\cite{Sch13} (with $t_0\approx{}2.5\,$eV the equilibrium hopping for
nearest-neighbors) we predict the narrow Mott-insulating phase (MI),
corresponding the strain adjusted such that a~single-particle spectrum
is gapless, to be surrounded with band-insulating phases (BI).
A~significant gap reduction in MI compared to BI allows to expect
the former to behave almost as a~semimetallic system in the presence
of external bias voltage or thermal excitations. 

The remaining part of the paper is organized as follows.
In Section~\ref{modmet}
we briefly present the effective Hubbard Hamiltonian and the Gutzwiller
Approximation (GA). Also in Section~\ref{modmet}, a~simplified relation
between the model parameters and geometric strains is put forward. 
In Section~\ref{resdis}, we discuss our numerical results concerning
the phase diagram of the effective Hubbard Hamiltonian adapted to model
zigzag $(10,0)$ nanotube subjected to the longitudinal strain, 
as well as the strain effect on multiparticle charge-energy gap. 
The conclusions are given in Section~\ref{conclu}.

\section{Model and methods}
\label{modmet}

\begin{figure*}[!t]
  \includegraphics[width=\linewidth]{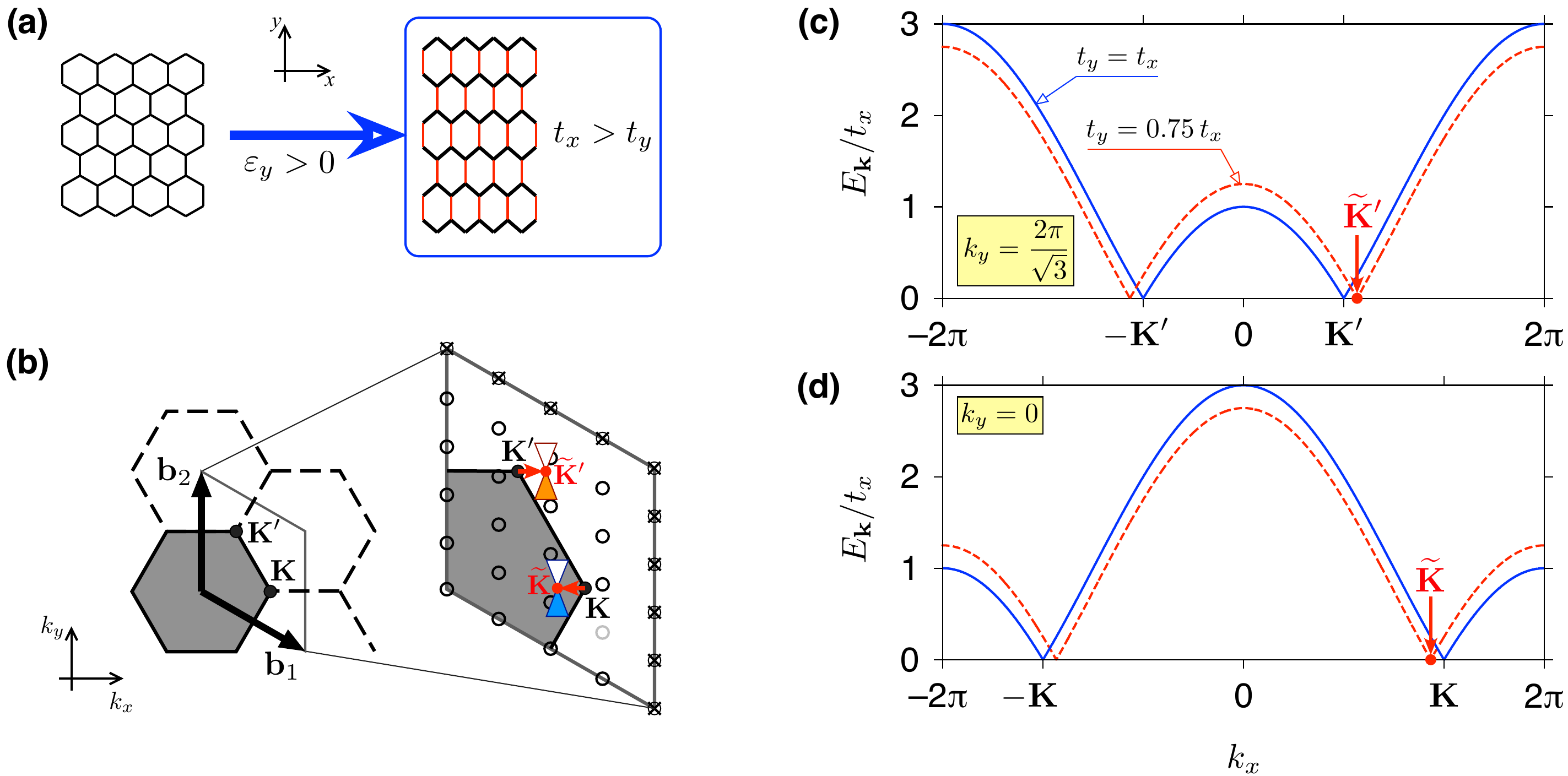}
  \caption{ \label{kkshifts:fig}
    (a) Honeycomb lattice subjected to uniaxial strain in the armchair
    direction. (b) Hexagonal first Brillouin zone (FBZ) of the reciprocal
    lattice, with (dimensionless) basis vectors
    ${\bf b}_1=\left(2\pi/\sqrt{3}\right)\,(\sqrt{3},-1)$ and
    ${\bf b_2}=\left(2\pi/\sqrt{3}\right)\,(0,2)$.
    The magnified area shows discretized FBZ for a finite system of
    $N = 2N_xN_y$ atoms with periodic boundary conditions [see Eq.\
    (\ref{kxynxy})] with the Dirac points shifted the positions
    $\mbox{{\boldmath$\tilde{K}$}}$ and $\mbox{{\boldmath$\tilde{K}'$}}$
    due to the strain. 
    (c) Cross section of the bulk single-particle energy [see Eqs.\
    (\ref{epsilonk}) and (\ref{akbkdef})] for $k_y=2\pi/\sqrt{3}$, matching
    the upper edge of FBZ.
    Blue solid line correspond to the lattice in equilibrium ($t_x=t_y$),
    red dashed is for the strain quantified by $t_y=0.75\,t_x$. 
    (d) Same as (c), but for $k_y=0$. 
  }
\end{figure*}

We write the Hamiltonian of the Hubbard model on anisotropic honeycomb
lattice as $H=H_t+H_U$ \cite{Per09,Rut23}, where the kinetic part is
\begin{align}
  H_t =
  &-t_x\sum_{\langle{}ij\rangle_x,s}(e^{{\rm i}\phi}c_{i,s}^{\dagger}c_{j,s}+\mbox{H.c.})
  \nonumber \\
  &-t_y\sum_{\langle{}ij\rangle_y,s}(c_{i,s}^{\dagger}c_{j,s}+\mbox{H.c.})
  \label{hamt}
\end{align}
and the interaction part is
\begin{equation}
  \label{hamu}
  H_U = U\sum_j{}n_{j\uparrow}n_{j\downarrow}. 
\end{equation}
Here, $c_{i,s}^{\dagger}$ ($c_{i,s}$) creates (annihilates) an electron at site
$i$ of the honeycomb lattice with spin $s=\uparrow$ or $\downarrow$, and
$n_{is}=c_{i,s}^{\dagger}c_{i,s}$ is the particle number operator. $t_x$ ($t_y$)
is the nearest-neighbor hopping along (out of) the zigzag direction.
The Peierls phase $\phi=\Phi/N_x$ with $\Phi$ being the magnetic flux
parallel to $y$ axis in a~zigzag $(N_x,0)$ geometry
(see Fig.\ \ref{setup-ab:fig}).
In the interaction part, $U$ is the on-site Hubbard repulsion and operator
$n_{j\uparrow}n_{j\downarrow}$
measures the number of double occupancies.
Throughout the paper, we focus on the ground-state phase diagram at half
filling, for $\Phi=0$ and $t_y\leqslant{}t_x$ (i.e., strain applied in
the armchair direction).

Several approximate methods can be utilized to determine whether the
ground state of the Hamiltonian $H$ is conducting or insulating.
Within the Gutzwiller Approximation (GA) \cite{Jed10,Che17} one needs
to find the minimum of
\begin{align}
  \frac{E_G^{\rm (GA)}}{N} &= q(m,d)\times \nonumber \\
  &
  \left[
    -\frac{2}{N}\sum_{\bf k}\sqrt{E_{\bf k}^2+\left(\frac{Um}{2}\right)^2}
    + \frac{Um^2}{2}
  \right] + Ud, 
  \label{eggamd}
\end{align} 
with respect to the sublattice magnetization $m$ (quantifying
the antiferromagnetic order) and the average double occupancy $d$. 
The constrictions are $|m|\leqslant{}1$ and
$d\leqslant{}d_{\rm max}=\frac{1}{4}(1-m^2)$. 
The band narrowing factor is given by 
\begin{equation}
  q(m,d) = 
   \frac{4d}{1\!-\!m^2}\left[
    1\!-\!2d+\sqrt{(1\!-\!2d)^2-m^2} 
  \right].
  \label{qumd}
\end{equation}
The summation in Eq.\ (\ref{eggamd}) runs over quasimomenta
${\bf k}\equiv{}(k_x,k_y)$ in the first Brillouin zone
(see Fig.\ \ref{kkshifts:fig}), namely
\begin{align}
  k_x &= \frac{2\pi{}}{N_x}(n_x+\phi),
  \ \ \ \ \ \ 
  k_y = \frac{4\pi{}}{\sqrt{3}}\left(\frac{n_y}{N_y}-\frac{n_x}{2N_x}\right),
  \label{kxynxy}
  \\
  n_x &= 0,1,\dots,N_x\!-\!1,  \ \ \ \ n_y = 0,1,\dots,N_y\!-\!1,
  \nonumber
\end{align}
with $N_{x,y}$ being the number of unit cells in $x,y$ direction, and
$N=2N_xN_y$ (the periodic boundary conditions are imposed). 
We further notice that setting $d=d_{\rm max}$ in Eqs.\ (\ref{eggamd}) and
(\ref{qumd}) reduces the procedure to the familiar Hartree-Fock approximation
(HF) with the antiferromagnetic order at half filling,
$\langle{}n_{i\uparrow}\rangle=\frac{1}{2}(1\pm{}m)$ and
$\langle{}n_{i\downarrow}\rangle=\frac{1}{2}(1\mp{}m)$, where
the upper (lower) sign corresponds to the sublattice $A$ ($B$). 

The single-particle energies for anisotropic honeycomb lattice are given by
\begin{equation}
  \label{epsilonk}
  E_{\bf k} = t_x\sqrt{a_{\bf k}^2+b_{\bf k}^2}, 
\end{equation}
with
\begin{align}
  a_{\bf k} =&
    -\cos\left(\frac{k_x}{2}+\frac{\sqrt{3}k_y}{2}\right)
    -\cos\left(\frac{k_x}{2}-\frac{\sqrt{3}k_y}{2}\right) - \frac{t_y}{t_x},
   \nonumber \\
  b_{\bf k} =& 
    \sin\left(\frac{k_x}{2}+\frac{\sqrt{3}k_y}{2}\right)
    -\sin\left(\frac{k_x}{2}-\frac{\sqrt{3}k_y}{2}\right).
    \label{akbkdef}
\end{align}

Also in Fig.\ \ref{kkshifts:fig}, we display cross sections of $E_{\bf k}$ for
$k_y=0$ and $k_y=2\pi/\sqrt{3}$ in the bulk limit
(i.e., $N_x,N_y\rightarrow{}\infty$).
In the vicinity of Dirac points, conical
band structure gets shifted along the $k_x$ axis by the value of
strain-induced vector potential, for $|t_x-t_y|\ll{}t_x$, 
\begin{equation}
  \label{delaxapp}
  \delta{}A_x \approx{} \mp{} \frac{2}{\sqrt{3}}\frac{t_x-t_y}{t_x}
  \approx{} \mp \frac{\sqrt{3}}{2}\beta{}\varepsilon_y(1+\nu), 
\end{equation}
where the upper (lower) sign correspond to $K$ ($K'$) valley,
$\beta\approx{}2-3$ is
the dimensionless electron-phonon coupling, $\varepsilon_y$ is relative strain
along the armchair direction, and $\nu=-\varepsilon_x/\varepsilon_y
\approx{}0.2-0.4$ is Poisson's ratio \cite{Gan21,strainfoo}.
Deriving the second approximate equality in Eq.\ (\ref{delaxapp}) we have
parametrized the hopping elements according to \cite{Dre98,Ryc13}
\begin{equation}
\label{tijdij}
  t_{x,y} = -t_0\left(1-\beta\frac{\delta{}d_{x,y}}{d_0}\right), 
\end{equation}
where the relative bond elongations, i.e., $\delta{}d_{x,y}\equiv{}d_{x,y}-d_0$
with the equilibrium bond length $d_0$, can be written as
\begin{equation}
  \frac{\delta{}d_x}{d_0} = \frac{1}{4}\varepsilon_y\left(1-3\nu\right)
    + {\cal O}(\varepsilon_y^2),
    \ \ \ \ \ \ 
  \frac{\delta{}d_y}{d_0} = \varepsilon_y.
\end{equation}

For a~zigzag $(N_x,0)$ geometry, $k_x$ gets quantized according to Eq.\
(\ref{kxynxy}) (hereinafter, we set $\phi=0$) and $k_y$ remains continuous
due to $N_y\rightarrow\infty$. In turn, the appearance of gapless subbands
is expected for
\begin{equation}
  N_x\left( \frac{1}{3} + \frac{\delta{}A_x}{2\pi}\right) = m,
  \ \ \ \text{with}\ \ \ m\ \ \text{integer}. 
\end{equation}
In the absence of strain, $\delta{}A_x=0$, the above reduces to 
$N_x=3m$, restoring the standard condition for metallicity of carbon nanotubes
with zigzag edges \cite{Cha07}. 
In Table~\ref{zerogaptytx}, we go beyond the small-strain limit of Eq.\
(\ref{delaxapp}), and present the values of $t_y/t_x$ corresponding to
vanishing single-particle gap, 
\begin{equation}
  \label{deltba}
  \Delta{}E_{\rm TBA} \equiv 2\min_{n_x,k_y} E_{\bf k} =0,
\end{equation}
which were obtained numerically for $10\leqslant{}N_x\leqslant{}20$. 
The case of $N_x=10$, showing only one zero for $t_y/t_x=0.618$
(corresponding to $\varepsilon_y\approx{}0.15$),
is chosen for further considerations.

\begin{table}[!hb]
\caption{
  Values of the hopping ratio corresponding to the vanishing single
  particle gap ($\Delta{}E_{\rm TBA}$) obtained from the anisotropic
  tight-binding Hamiltonian for small zigzag $(N_x,0)$ nanotubes.
  Notice that metallic nanotubes ($N_x=3m$) show
  $\Delta{}E_{\rm TBA}=0$ for $t_x=t_y$.
  \label{zerogaptytx}
}
\begin{tabular}{cccc}
\hline\hline
  $N_x$  &  \multicolumn{2}{c}{$(t_y/t_x)_{\Delta{}E_{\rm TBA}=0}$} & \\ \hline 
  $\ \ \ $10$\ \ \ $  & 0.6180 & & \\
   11  &  $\ \ \ $0.8308$\ \ \ $  &  $\ \ \ $0.2846$\ \ \ $  &  \\
   12  &  1.0000  &  0.5176  &  \\
   13  &  0.7092  &  0.2411  &  \\
   14  &  0.8678  &  0.4450  &  \\
   15  &  1.0000  &  0.6180  &  $\ \ \ $0.2091$\ \ \ $  \\
   16  &  0.7654  &  0.3902  &  \\
   17  &  0.8915  &  0.5473  &  0.1845  \\
   18  &  1.0000  &  0.6840  &  0.3473  \\
   19  &  0.8034  &  0.4910  &  0.1652  \\
   20  &  0.9080  &  0.6180  &  0.3129  \\
\hline\hline
\end{tabular}
\end{table}

\section{Results and discussion}
\label{resdis}

\begin{figure*}[!t]
  \includegraphics[width=\linewidth]{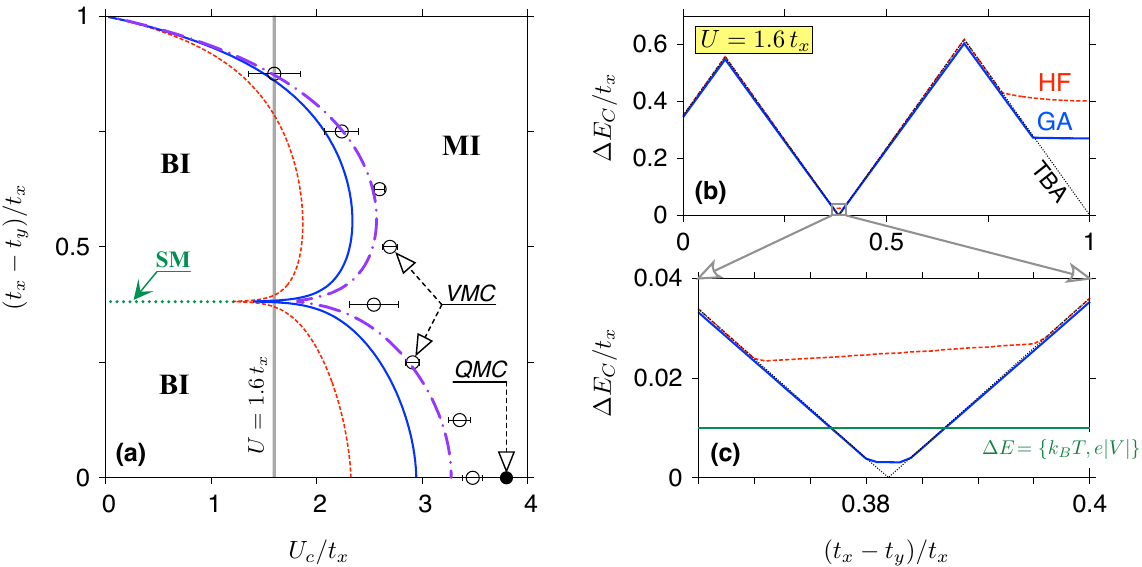}
  \caption{ \label{phdiag3pan:fig}
    (a) Phase diagram for the Hubbard model on anisotropic honeycomb lattice
    with $t_y\leqslant{}t_y$ [see Eqs.\ (\ref{hamt}) and (\ref{hamu})] and
    the number of unit cell along the zigzag direction $N_x=10$.
    Lines depict the critical Hubbard
    repulsion obtained within the Hartree–Fock method [red dashed], the
    Gutzwiller Approximation [blue solid], and Statistically-consistent GA
    [purple dash-dot]. Datapoints with errorbars are the results of VMC
    simulations for the Gutzwiller Wave Function reprinted from
    Ref.\ \cite{Rut23}.
    Quantum Monte Carlo value for the isotropic case ($t_x=t_y$),
    and the limit of $N_x,N_y\rightarrow{}\infty$ taken numerically,
    $U_c/t_0=3.86$ (see Ref.\ \cite{Sor12}), is marked with full circle.
    Remaining labels are: the 
    band insulator (BI), the Mott insulator (MI), and the semimetal (SM) 
    [mark with green dotted line corresponding to vanishing single-particle
    gap at $t_y/t_x=0.618$, see Table~\ref{zerogaptytx}].
    Grey vertical line depicts the value of $U/t_x=1.6$ used in remaining
    plots.
    (b) Charge gap as a~function of strain for $U/t_x=1.6$. Different
    lines correspond to the methods specified on the plot. 
    (c) Zoom-in of (b) for the vicinity of 
    $t_y/t_x=0.618$.
    Green horizontal line at $\Delta{}E=0.01\,t_x\approx{}25\,$meV marks
    typical energy of quasiparticle excitations at $T=300\,$K.
    (We further notice that a~comparable energy of $e|V|$ may also appear
    due to source-drain voltage difference.) 
  }
\end{figure*}

It is clear from Eq.\ (\ref{delaxapp}) that dimensionless parameter
$(t_x-t_y)/t_x\propto{}\varepsilon_y$ may be used to quantify uniaxial
strain along the armchair direction. To quantify the Hubbard interaction,
we choose $U/t_x$. (It can be shown that
$t_x\approx{}t_0[1-\frac{1}{4}\beta{}\varepsilon_y(1-3\nu)]$; therefore,
for $\nu\approx{}1/3$ the strain dependence is weak.) 

Our central results are presented in Fig.\ \ref{phdiag3pan:fig}, where we
display the phase diagram for the Hamiltonian $H=H_t+H_U$, see Eqs.\
(\ref{hamt}) and (\ref{hamu}), and $N_x=10$.
The phases are identified by finding the optimal parameters $(m,d)$ within
GA. Blue solid line in panel (a) marks the border between the solutions
with $m=0$ and $m\neq{}0$ (i.e, the critical value of $U_c^{\rm (GA)}$).
Similarly, by fixing the second parameter at $d=d_{\rm max}$
[see Eq.\ (\ref{eggamd})], we determine the value of $U_c^{\rm (HF)}$
(red dashed line). Purple dashed-dotted line marks the value of
$U_c^{\rm (SGA)}$ obtained within the {\em Statistically-consistent GA}
\cite{Jed10},
a~method in which not only parameters $(m,d)$ but
also hopping integrals of the auxiliary Hamiltonian are
optimized. It can be noticed that for $(t_x-t_y)/t_x\rightarrow{}1$, all
the methods leads to $U_c\rightarrow{}0$, reproducing the value for
decoupled Hubbard chains \cite{Lie03}.

The solution with $m\neq{}0$ is interpreted as the Mott insulator (MI).
If $m=0$, the interpretation depends on whether a~single-particle gap
$\Delta{}E_{\rm TBA}=0$ or $\Delta{}E_{\rm TBA}>0$.
In the former case, occurring only for $t_y/t_x=0.618$ (green dotted line)
we recognize the semi-metallic phase (SM). Otherwise, the ground state
can be identified as the band insulator (BI). 

Unlike for bulk graphene, for which $U_c$ monotonically decreases with
$(t_x-t_y)/t_x$ \cite{Rut23}, for $N_x=10$ we observe an anomaly near
$t_y/t_x=0.618$ corresponding to $\Delta{}E_{\rm TBA}=0$.
The reduction of $U_c$ can be attributed to the appearance of a~nonzero
density of states at the Fermi level, that could produce magnetic
instability for (in principle) any $U>0$. 
However, the value of $U_c\approx{}0$ (away from $t_y/t_x=0$) is not
supported  with our results. A~series of approximations applied leads to
$0<U_c^{\rm (HF)}<U_c^{\rm (GA)}<U_c^{\rm (SGA)}$ for any $0<t_y/t_x\leqslant{}1$. 
In particular, for $t_y/t_x=0.618$ we obtain
$U_c^{\rm (HF)}=1.21\,t_x$, $U_c^{\rm (GA)}=1.39\,t_x$, and
$U_c^{\rm (SGA)}=1.82\,t_x$, each showing a~significant (approximately
a~$40$ percent) reduction in comparison to the bulk case.
Probably, the competition of different insulating ground states, combined
with quantum fluctuations, leads to the instability resulting in the
appearance of a~conducting phase, in a~similar manner as earlier discussed
on the examples of one-dimensional correlated nanosystems
\cite{Gan20,Ryc01,Spa01,Kad15}. 

We further notice that the Variational Monte Carlo (VMC) results for
$N_x=N_y=10$, reprinted from Ref.\ \cite{Rut23} (datapoints with errorbars),
are very close to the SGA results, showing that the latter can be regarded
as a~computationally-inexpensive counterpart to the former. 

Next, in Figs.\ \ref{phdiag3pan:fig}(b) and \ref{phdiag3pan:fig}(c),
we have fixed the Hubbard interaction at $U=1.6\,t_x$ \cite{Sch13}, and 
displayed the charge-energy gap for a~correlated state,
\begin{align}
  \Delta{}E_C &= E_G^{(N_e=N\!+\!1)} + E_G^{(N_e=N\!-\!1)} - 2E_G^{(N_e=N)}
  \nonumber \\
  & \approx{} 2q(m,d) \min_{n_x,k_y}\sqrt{E_{\bf k}+(Um/2)^2}, 
\end{align}
with $N_e$ denoting the number of electrons and
$q(m,d)$ given by Eq.\ (\ref{qumd}), as a~function of $(t_x-t_y)/t_x$.
Here, the parameters $(m,d)$ are optimized only once, for half filling
($N_e=N$); HF corresponds to a~fixed $d=d_{\rm max}$; the gap following from
Tight-Binding Approximation (TBA) is given by the first equality
in Eq.\ (\ref{deltba}).

Although a~simple case of $N_x=10$, showing the only one (nontrivial) zero
of $\Delta{}E_{\rm TBA}$, is considered, quasiperiodic behavior of $\Delta{}E_C$
with increasing strain is already visible in Fig.\ \ref{phdiag3pan:fig}(b).
A~zoom-in of the area surrounding $\Delta{}E_{\rm TBA}=0$, presented in
Fig.\ \ref{phdiag3pan:fig}(c), shows that the multiparticle gap
($\Delta{}E_C$) is reduced, comparing to the equilibrium value of
$\Delta{}E_C\approx{}0.35\,t_x$ for $t_y=t_x$, by a~factor varying
(depending on the method) from $\sim{}10$ (HF) to $\sim{}100$ (GA) when
$t_y/t_x\approx{}0.618$.
(Notice that $U_c^{\rm (SGA)}>1.6\,t_x$ for such a~range, and the
semimetallic phase is predicted within SGA.)
$\Delta{}E_C$ also shows two cusp-shaped maxima, located near $t_y/t_x=0.309$
and $0.897$, for which nearest values of $k_x$ are equally distant from
the Dirac point. In such cases, the role of electron correlations is
negligible due to closed subbands (notice that the the lines corresponding
to different approximations overlap). 

Significantly, a~narrow gap following from GA for $t_y/t_x\approx{}0.618$
ensures that $\Delta{}E_C$ becomes negligible in the presence of thermal
excitations characterized by the energy of, say $\Delta{}E\approx{}t_0/100
\approx{}25\,$meV (at $T=300\,$K), or a~comparable energy of $e|V|$
due to source-drain voltage difference in transport experiment.
In turn, for $U\approx{}1.6\,t_x$ and $t_y/t_x\approx{}0.618$,
the semimetallic (or almost-semimetallic) behavior is predicted.

\section{Conclusions}
\label{conclu}

We have investigated, using the Gutzwiller approximation and related methods,
the mutual effect of strain and electron-electron interaction on the
ground state of semiconducting nanotube with zigzag edges.
The results suggest that approximately $15\%$ strain along the main axis
may drive $(10,0)$ nanotube into the semimetallic phase.
For weak Hubbard repulsion, we expect the semimetallic phase to re-appear
generically, for strains and winding numbers adjusted such that
a~single-particle gap vanishes. 
Most remarkably, quasiperiodic bandgap oscillations with the increasing
strain are predicted. 
The analogy to the Aharonov-Bohm effect at nanoscale is put forward.

\section*{Acknowledgments} 
We thank J\'{o}zef Spa\l{}ek for discussions.
Support from the National Science Centre of Poland (NCN), via Grant
No.\ 2014/14/E/ST3/00256, at the early stage is acknowledged. 
Computations were partly performed using the PL-Grid infrastructure.




\begin{thebibliography}{}
\bibitem{metalfoo}
More detailed modelling suggests that the ground state of single
wall nanotube zigzag edges may evolve from moderate- to narrow-gap
semiconductor, the metallic phase is possible for armchair nanotubes;
see Ref.\ \cite{Shy21}.
\end{thebibliography}
\end{document}